\documentclass[11pt]{article}

\usepackage[margin=1in]{geometry}
\usepackage[english]{babel}

\usepackage{amsmath, amssymb}
\usepackage{graphicx}
\usepackage{booktabs}
\usepackage{multirow}
\usepackage{tabularx}
\usepackage{float}
\usepackage{subfig}
\usepackage{enumitem}
\usepackage{comment}
\usepackage{tcolorbox}
\usepackage{listings}
\usepackage{xcolor}
\usepackage{orcidlink}
 \usepackage{natbib} 

\usepackage{hyperref}
\hypersetup{
    colorlinks=true,
    urlcolor=blue,
    linkcolor=blue,
    citecolor=black
}

\title{Assessing the Geographic Diversity of AI's Platial Representations in Image Generation}

\author{
Zilong Liu \and Krzysztof Janowicz \and Mina Karimi \\
\\
Department of Geography and Regional Research, University of Vienna, Austria
}

\date{}

\begin{document}

\maketitle

\begin{abstract}
(Gen)AI diversity is not merely an ethical issue. From the perspective of geographic information science (GIScience), it could be interpreted as a function of uncertainty and as a form of cognitive bias, embedded in AI outputs. Recent work has sought to develop information-theoretic diversity measures and apply them to evaluate AI-chatbot outputs in a geographic context. As the AI ecosystem to which we are exposed on a daily basis becomes rapidly multimodal, we believe it is important to examine \textit{geographic diversity} across various modalities. Focusing on images, this paper aims to fill this research gap. First, we select the GPT and DALL·E models as state-of-the-art examples and point out how assessing their geographic diversity involves various stages, including prompt revision and image generation. Then, taking inspiration from species diversity measures in ecological research, we incorporate similarity weighting into the measurement of geographic diversity. Next, we demonstrate how to evaluate geographic diversity in image generation through a case study. Our analysis reveals several counterintuitive findings. For instance, older models can exhibit greater geographic diversity despite producing lower-quality images, and prompt revision yields greater geographic diversity than image generation. At the same time, we observe explicit model homogeneity underlying the lack of geographic diversity, as the selected models consistently depict the same prototypical geo-specific feature or similar features. This is concerning, as it risks producing stereotypical representations of places.
\end{abstract}

\noindent\textbf{Keywords:} Geographic Diversity; Hill Number; Leinster-Cobbold Number; Image Generation; AI Representation

\section{Introduction}
\label{sec:intro}
Modern (Gen)AI models rank among the most advanced AI systems in operation today. Despite their capabilities, they are known to produce outputs lacking diversity. To give just a few examples, when asked to guess a number between 1 and 50, popular models disproportionately respond with the number \textit{27}~\citep{faraaz2025llms}; when prompted to picture a person, they frequently depict \textit{male} figures~\citep{naik2023social}; and when requested to name a country, most of them default to common and invariant examples, such as \textit{Canada} \citep{https://doi.org/10.1111/tgis.70242}. Given that these models are also foundation models~\citep{bommasani2021opportunities}, their bias, i.e., \textit{regime of representation}~\citep{qadri2023ai}, hinders their intended, \textit{general-purpose} utility. It constrains their production of \textit{pluralistic} outputs that align with the variability of human expectations~\citep{sorensen2024roadmap,janowicz2025whose}.

Why should GIScientists worry about such diversity issues? Here, we raise two points centered around geographic information, which remains relevant well beyond model training (i.e., an aspect already embraced by the \textit{GeoAI} community~\citep{janowicz2020geoai}). The first aspect concerns the \textit{uncertainty} inherent in geographic information~\citep{couclelis2003certainty}. Diversity can be interpreted as a function of the uncertainty in how geographic information is encoded, transmitted, and decoded by an AI model, reflecting its \textit{stochastic} nature~\citep{bender2021dangers}. The second point concerns the \textit{cognition} of geographic information~\citep{montello2004cognition}. A lack of diversity can be interpreted as an AI model's fallacy to rely on a fixed geographic category structure anchored in \textit{prototypical} instances, indicating its cognitive biases~\citep{rudolph2025bias}.

Most recently, research has begun to measure diversity in a geographic context in order to evaluate how representative (Gen)AI outputs are. \citet{liu2025operationalizing} found certain countries and continents more frequently elicited from AI chatbots, specifically autoregressive large language models (LLMs) with question-answering capabilities. Based on this phenomenon, they defined \textit{geographic diversity} in terms of two aspects: \textbf{(1)} the distinct number of places and \textbf{(2)} the balance of their sampling distribution during the generation process. Stemming from Shannon entropy~\citep{shannon1948mathematical}, which quantifies the average uncertainty of information content, they further developed diversity measures to capture the richness \textit{and} the evenness of places referenced in model outputs. Such effort reflects a meaningful theme: combining probabilistic modeling and information theory to provide measurements about the (im)balanced representation of geographic entities and phenomena.

However, to date, the operationalization of geographic diversity has been documented for text generation alone. This leaves corresponding work on AI image generation unaddressed. Unlike text output, which can be reduced to the level of individual tokens, AI-generated images contain rich information equivalent to thousands of tokens. For instance, an image generated by Stable Diffusion~\citep{rombach2021highresolution}, a representative text-to-image (T2I) generation model, consists of at least 512$\times$512 pixels. It further requires human knowledge, or even another trained AI system, to semantically segment such images and detect their constituent objects.

In this paper, we aim to address the research gap of geographic diversity in AI's image generation by offering both methodological insights (e.g., how to measure and interpret) and empirical observations (e.g., patterns in model outputs). \textbf{Our research contributions are as follows}.
\begin{itemize}
    \item Using models from the DALL·E and GPT families, we show that state-of-the-art proprietary image generation relies on a multi-agent system. In this setup, an LLM first revises user prompts and then calls a T2I model with the revised prompts~\citep{openai_image_generation_guide}. Hence, we highlight that the evaluation of geographic diversity requires accounting for various stages and addressing \textbf{(Gen)AI's multimodal nature}.
    \item We demonstrate that developing more holistic measures of geographic diversity benefits from accounting for similarity in a geographic context. Analogous to prior work in ecological research that incorporates similarity weights into species diversity~\citep{leinster2012measuring}, we propose \textbf{similarity-sensitive measures for geographic diversity}. We further demonstrate how to use auxiliary knowledge bases to obtain categorical information to support relevant similarity computations.
    \item We conduct a case study on the city of Vienna, which is actively adopting AI image generation for participatory urban planning in Austria~\citep{dialogplus2025_ki_partizipation}. We discover prototypical geo-specific features (e.g., landmarks) emerging consistently from model outputs across the generative pipeline. By profiling both similarity-insensitive and similarity-sensitive geographic diversity, we further observe \textbf{an explicit and shared lack of diversity} throughout the pipeline.
\end{itemize}

In doing so, we deepen the understanding of geographic diversity in GenAI, showing that it concerns not only the naming of places (e.g., on a coarser geographic scale) but also the depiction of a specific place (i.e., on a finer geographic scale). Accordingly, our work adds the \textbf{diversity of AI's platial representations} as a new dimension of geographic diversity, strengthening the role of \textit{place} as a first-class component of understanding and quantifying biases in generative (Geo)AI systems.

The remainder of this paper is organized as follows. Section~\ref{sec:background} provides our research background. Section~\ref{sec:methods} describes our methods. Section~\ref{sec:results} presents our results. Section~\ref{sec:discussion} includes our discussion. Section~\ref{sec:conclusions} concludes this paper.

\section{Background}
\label{sec:background}
In this section, we review background literature on \textbf{(1)} information theory, \textbf{(2)} species diversity, and \textbf{(3)} geographic diversity. Both the measurement of species diversity and geographic diversity have roots in information theory. In addition, the current operationalization of geographic diversity was built directly upon species diversity.

\subsection{Information Theory}
In a nutshell, information theory is the formal study of the coding and transmission of information. It originated from the work of \citet{shannon1948mathematical}, who introduced (Shannon) entropy as the quantification of the expected amount of information associated with a discrete random variable $X$. Such quantification is equivalent to a measure of the average uncertainty of the variable's possible outcomes. In Eq.\,(\ref{eq:entropy}), $x$ represents a possible outcome randomly drawn from the set $X$ and $p(x)$ represents the probability of observing $x$.

\begin{equation}
H(X) = - \sum_{x \in X} p(x) \log p(x)
\label{eq:entropy}
\end{equation}


Consider the weather in a hypothetical city, where each day can be sunny, cloudy, windy, rainy, stormy, or snowy, and each outcome occurs with an equal probability of $\frac{1}{6}$. The entropy of the weather is therefore $H(X)=\log(6)$, which is approximately 2.6 if the logarithm is taken with base 2. This means that knowing the weather of a random day conveys roughly 2.6 \textit{bits} of information.

\citet{shannon1948mathematical} also developed the concept of \textit{information content}, which quantifies the amount of information associated with a single outcome. Eq.\,(\ref{eq:selfinfo}) presents its formula. From the previous \textit{weather} example, one can derive that observing a sunny, cloudy, windy, rainy, stormy, or snowy day in our hypothetical city also conveys roughly 2.6 bits of information, since $I(x)=\log(6)$.

\begin{equation}
I(x) = -\log p(x)
\label{eq:selfinfo}
\end{equation}

\subsection{Species Diversity}
Historically, the bulk of the literature on diversity has concentrated on species diversity. Ecologists have treated diversity as one of the most important aspects of ecological communities, and they view its measurement as crucial for understanding, for instance, the resilience of a community in the face of anthropogenic impacts. It is widely acknowledged that species diversity of an ecological community comprises two fundamental components: \textit{species richness} and \textit{species evenness}. Species richness refers to the number of distinct species in an ecological community, and species evenness refers to their relative abundance (i.e., the distribution of individual counts among species).

A standard measure of species diversity is the \textit{Hill number}~\citep{hill1973diversity}, also referred to as the \textit{equivalent number of species}. Eq.\,(\ref{eq:hill}) presents its formula, where $q$ denotes the order of the Hill number. It controls the sensitivity of the Hill number to rare versus common species. Here,  $X$ represents the species composition of the community, and $x$ represents a species randomly selected from the community.

\begin{equation}
{}^q\!D = \left( \sum_{x \in X} p(x)^q \right)^{\frac{1}{1-q}}
\label{eq:hill}
\end{equation}

Suppose there are two forests, each containing $1,000$ trees. Forest A has 10 different species: each represented by 100 trees. Forest B has 2 species: 900 trees of species $\text{B}_1$ and 100 tree of species $\text{B}_2$. When $q=0$, species diversity reduces to species richness, resulting in ${}^0\!D$ of 10 for Forest A and 2 for Forest B. When $q=2$, species diversity is weighted heavily towards species evenness, resulting in ${}^2\!D$ of 10 for Forest A and approximately 1.2 for Forest B. Here, the order-2 Hill number ${}^2\!D$ is also known as the \textit{Inverse Simpson Index}~\citep{simpson1949measurement}. Across these example orders $q$, it is clear that Forest A consistently exhibits higher diversity, and its diversity remains unchanged across orders, while Forest B’s diversity decreases as more emphasis is placed on evenness.

\citet{jost2006entropy} explained the relationship between entropy in information theory and diversity in ecological research. On the one hand, they clarified that entropy is not equivalent to diversity. For instance, it is ecologically intuitive that a community with $N$ equally abundant species has a diversity of $N$, but Shannon entropy would provide $log(N)$. On the other hand, they showed that entropy can be mathematically transformed into \textit{true} diversity. For instance, Shannon entropy (using the natural logarithm) can be exponentially transformed into ${}^1\!D$, i.e., the order-1 Hill number. Eq.\,(\ref{eq:transform}) presents this transformation. Therefore, from an information-theoretic perspective, measuring species diversity can be understood as quantifying the average uncertainty associated with observing a different species in an ecological community of interest.

\begin{equation}
{}^1\!D = \lim_{q \to 1} {}^q\!D = \exp\big(H(X)\big)
\label{eq:transform}
\end{equation}

\subsection{Geographic Diversity}
\citet{liu2025operationalizing} drew an analogy between \textit{species} in species diversity and \textit{place} in geographic diversity. They argued that, compared to \textit{space}, which is often modeled as a continuous random variable by GIScientists, \textit{place} can be modeled as a discrete random variable, which is suitable for probabilistic modeling. Also, they justified richness and evenness as the key dimensions of geographic diversity. Their justification is based on the work of \citet{shankar2017no}, who were among the first raising the issue of geographic diversity in benchmark datasets used for machine learning--based image classification, pointing out the lack of available data for certain countries.

As such, \citet{liu2025operationalizing} proposed applying the Hill number~\citep{hill1973diversity} to the measurement of geographic diversity, calling it the \textit{equivalent number of places}. In the current formulation of geographic diversity, $X$ represents the place composition of (Gen)AI outputs, and $x$ represents a place randomly drawn from (Gen)AI outputs.

They further built upon the work of \citet{lin1998information}, who noted that if a probabilistic model can be applied to a domain (e.g., \textit{similarity}), an \textit{information-theoretic definition} of that domain can be derived from information content. Based on these insights, and drawing on the work of \citet{jost2006entropy}, they suggested that entropy provides an information-theoretic definition of diversity both ecological and geographic contexts. Accordingly, measuring geographic diversity can be understood as quantifying the average uncertainty associated with observing a different place in the output of a (Gen)AI model of interest.

\begin{figure}[h!tbp]
    \centering
    \includegraphics[width=0.6\textwidth]{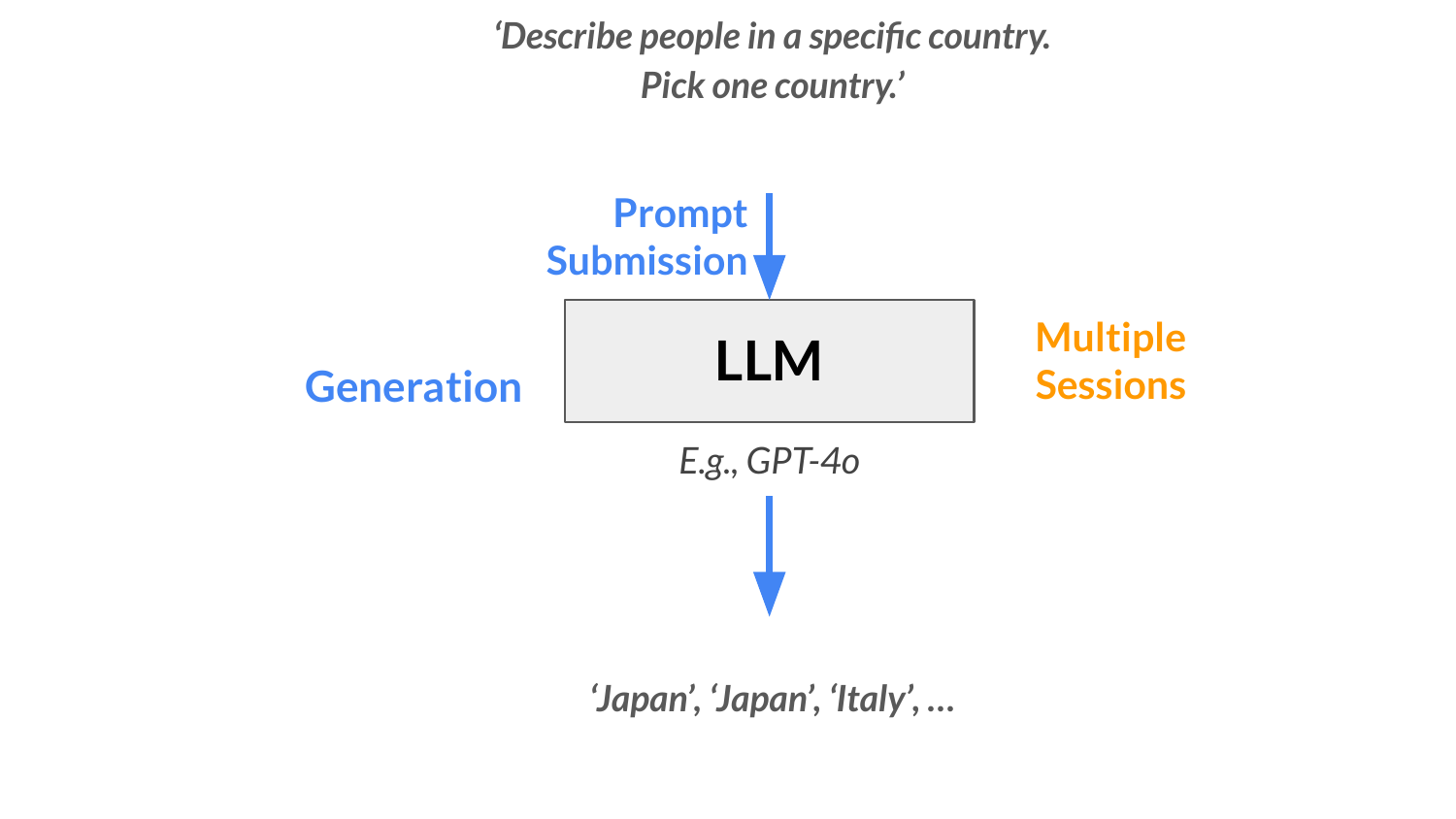}
    \caption{The experimental workflow used to test the geographic diversity of AI chatbots, adapted from \citet{liu2025operationalizing}.}
    \label{fig:place_naming}
\end{figure}

Later, \citet{liu2025operationalizing} applied their measure to the diversity evaluation of AI chatbots. By conducting multiple cold, independent sessions in which LLMs were prompted to name countries or continents, they obtained a collection of model outputs with rich place mentions. These outputs were subsequently used to evaluate geographic diversity based on the order-0, order-1 and order-2 Hill numbers, i.e., ${}^0\!D$, ${}^1\!D$, and ${}^2\!D$. Fig.\,\ref{fig:place_naming} illustrates their experimental workflow, which led to the discovery of not only a lack of geographic diversity but also a series of prototypical places, such as \textit{Japan} in response to the example user prompt. Their analysis also revealed that geographic diversity does not necessarily rise with the recency of the model.

\section{Methods}
\label{sec:methods}
In this section, we describe our experiment setup, including \textbf{(1)} image generation and \textbf{(2)} diversity measurement.

\subsection{Image Generation}


\citet{sumers2023cognitive} proposed augmenting LLMs with \textit{cognitive architectures}, such as in-context memory, reasoning, and tool use. These architectures can equip LLMs with contextual knowledge to interact with external environments, thereby transforming them to autonomous \textit{language agents}. OpenAI is a leading (Gen)AI company working in this direction, developing multimodal language agents alongside T2I models. This is the reason why we target our diversity evaluation at their models.

\subsubsection{OpenAI Models}
Among OpenAI's autoregressive LLMs, GPT-4o was the first to be equipped with tool-calling capabilities in image generation \citep{openai2025_image_generation}. It can be connected to GPT Image 1, the first OpenAI model supporting T2I generation via the Responses API. This API allows images to be generated through tool calls issued by OpenAI LLMs. Currently, GPT-4o and subsequent OpenAI LLMs can access GPT Image 1 or its mini variant, i.e., GPT Image 1 Mini, using this functionality~\citep{openai_image_generation_guide}. In addition, OpenAI Image API allows direct image generation from text prompts for all its T2I models. Besides GPT Image 1 and GPT Image 1 Mini, these models include GPT Image 1.5 and the DALL·E family of models. GPT Image 1.5 is the latest OpenAI T2I model at the time of writing, while the (older) DALL·E family includes DALL·E 2 and DALL·E 3.

\begin{figure}[h!tbp]
    \centering
    \includegraphics[width=0.6\textwidth]{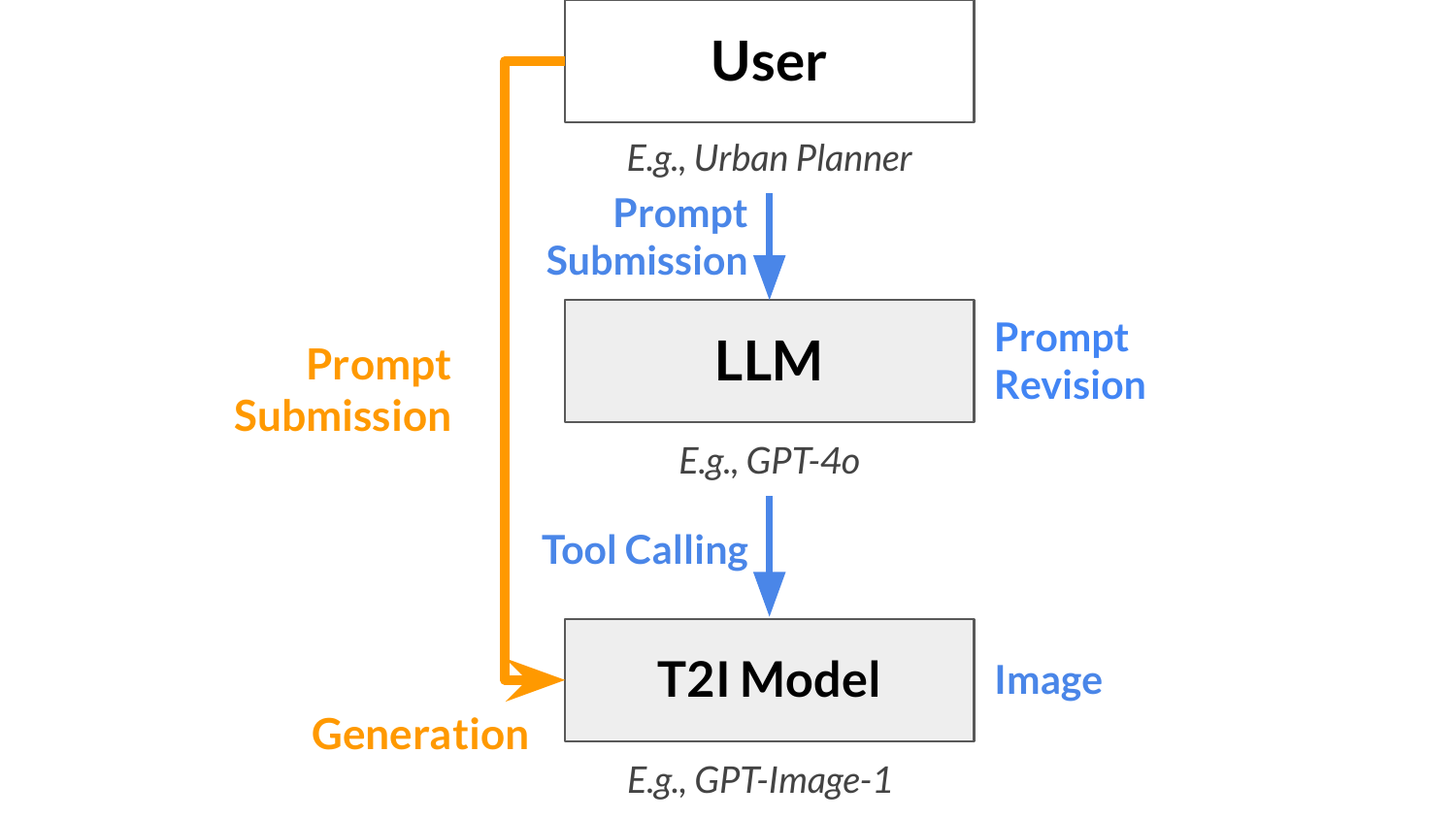}
    \caption{OpenAI's multi-agent system used for image generation, adapted from \citet{openai_image_generation_guide}. The automatic prompt revision performed by DALL·E 3~\citep{openai_dalle3_image_generation_guide} is not shown for simplicity.}
    \label{fig:multi_agent_pipeline}
\end{figure}

Fig.\,\ref{fig:multi_agent_pipeline} illustrates the differences between using a T2I model independently and in collaboration with an LLM for OpenAI's image generation. When a user submits a prompt, it can be passed directly to a T2I model for image generation. However, DALL·E 3 performs automatic prompt revision prior to image generation~\citep{openai_dalle3_image_generation_guide}. When a prompt is submitted to GPT-4o or it subsequent LLM, it will first be processed by the LLM, which also revises the prompt automatically before passing it to the GPT Image 1 (Mini) to generate the final image.

\subsubsection{Model Selection}
In our experiment, we aim to provide a focused case study using OpenAI models. Therefore, we include all OpenAI T2I models in our evaluation and select GPT-4o as our LLM of interest; see Table~\ref{tab:model} for details on the model snapshots we use.

\begin{table}
    \centering
    \caption{OpenAI models used in our experiment to achieve image generation. GPT-4o is an LLM working with GPT Image 1 or GPT Image 1 Mini to achieve image generation.}
    \begin{tabular}{lll}
        \hline
         \textbf{Model}& \textbf{Snapshot}  &\textbf{Type}  \\ \hline
         GPT Image 1.5& gpt-image-1.5-2025-12-16  &T2I  \\
         GPT Image 1& gpt-image-1  &T2I  \\
         GPT Image 1 Mini& gpt-image-1-mini &T2I  \\
         DALL·E 3& dall-e-3  &T2I  \\
         DALL·E 2& dall-e-2  &T2I  \\
         GPT-4o& gpt-4o-2024-08-06 &LLM  \\ \hline
    \end{tabular}
    \label{tab:model}
\end{table}

\subsubsection{Prompt Settings}
For our case study, we apply the same prompt(s) for both GPT-4o when calling GPT Image 1 (Mini) and to all T2I models independently. To ensure that our measurement is statistically significant, we follow the strategy of \citet{liu2025operationalizing} to conduct multiple sessions for each multi-agent system and for each T2I model independently. All other prompting parameters (e.g., quality option for a T2I model, or temperature and \textit{top\_p} for GPT-4o) are left at their default settings.

\noindent\textbf{System prompt:} \textit{[None]}

\noindent\textbf{User prompt:} \textit{Generate an image of Vienna.}

\noindent\textbf{Number of sessions:} 30 sessions per model

\subsubsection{Example Generation}
\begin{figure}[h!tbp]
    \centering
    \includegraphics[width=0.6\textwidth]{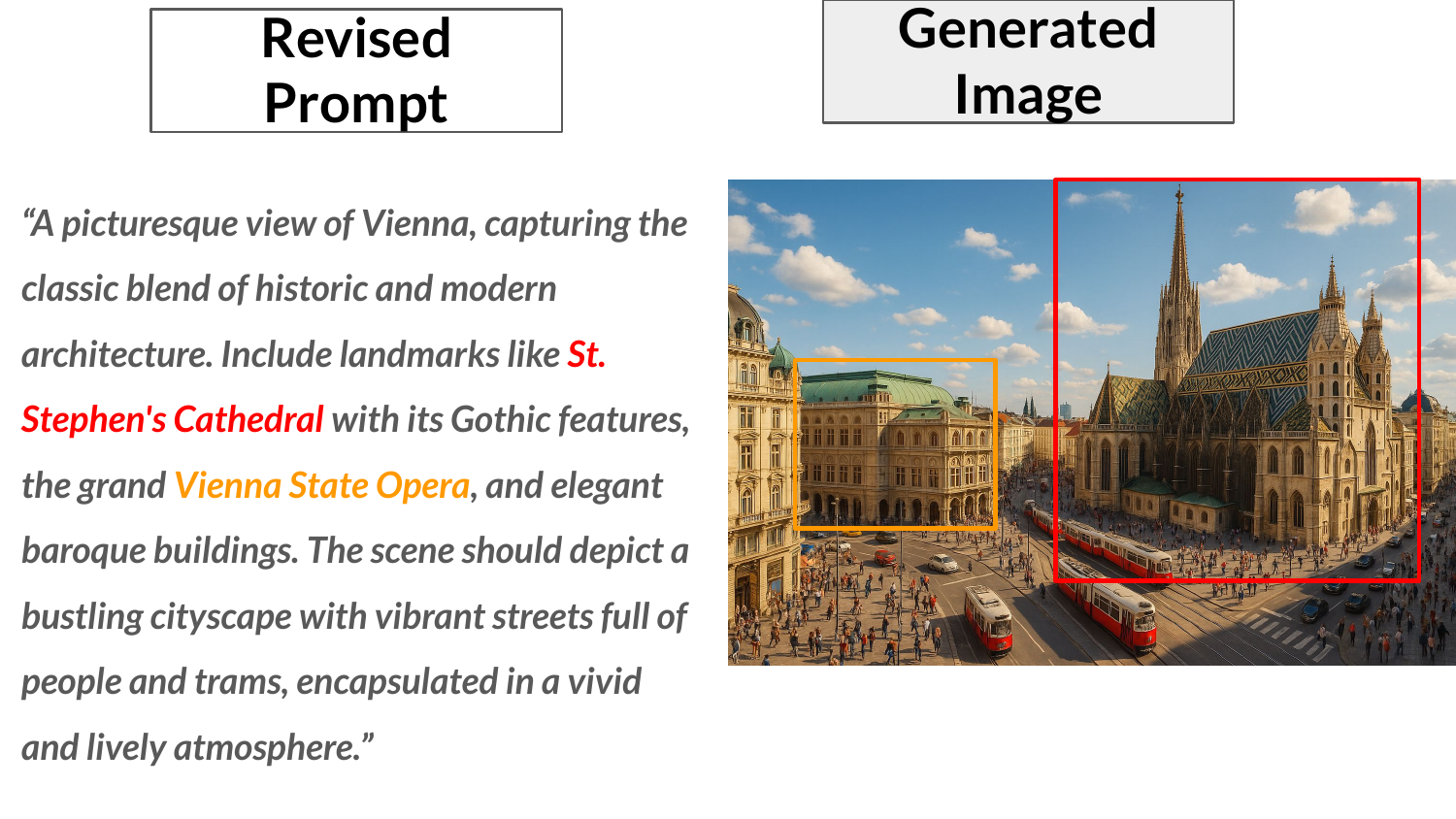}
    \caption{An example session where GPT-4o and GPT Image 1 (Mini) collaborate to generate images. The red text in the revised prompt and the red bounding box in the generated image refers to St. Stephen's Cathedral. The orange highlights the Vienna State Opera house.}
    \label{fig:multi_agent_example}
\end{figure}

Fig.\,\ref{fig:multi_agent_example} illustrates the revised prompt and the generated image produced by the multi-agent system, which is our primary focus, in one example session. The revised prompt mentions St. Stephen's Cathedral and the Vienna State Opera house, both of which are well-known landmarks in Vienna. From the generated image, St. Stephen's Cathedral appears in the foreground, whereas the Vienna State Opera house appears in the background.

\begin{figure}[h!tbp]
    \centering
    \includegraphics[width=0.6\textwidth]{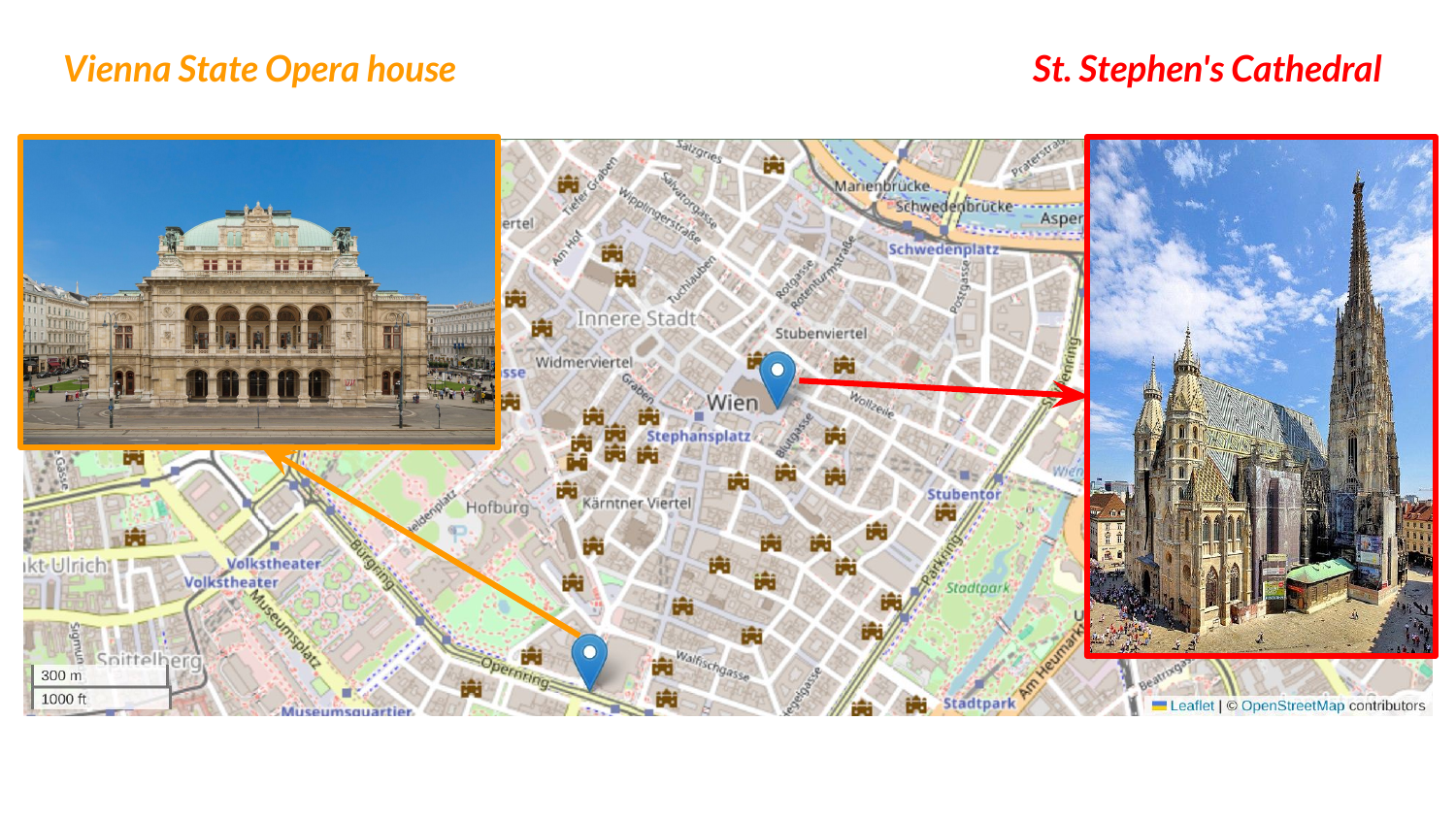}
    \caption{Real-world locations of St. Stephen’s Cathedral and the Vienna State Opera house, along with their real-world images.}
    \label{fig:reality}
\end{figure}

It is worth noting that, in reality, one cannot see both landmarks from this viewpoint. For reference, Fig.\,\ref{fig:reality} shows their real-world locations overlaid on OpenStreetMap\footnote{\url{https://www.openstreetmap.org}}, with images sourced from their English Wikipedia\footnote{\url{https://en.wikipedia.org}} pages.

\subsection{Diversity Measurement}
The current measure of geographic diversity originates from the measure of species diversity, and therefore, takes the same form, i.e., the Hill number~\citep{hill1973diversity}. However, \citet{leinster2012measuring} pointed out that the Hill number is a \textit{naive} model of diversity in the sense that it does not account for differences among species within an ecological community of interest. They argued that it is ecologically meaningful to state that a community with $N$ highly different species is more diverse than a community with $N$ highly similar species, even when their degrees of species evenness are identical.

\subsubsection{Place-Type Similarity}
Here, we argue that geographic diversity can also benefit from accounting for similarity. Since \textit{place} is the primary statistical unit of geographic diversity, we propose focusing on place similarity. In practice, place similarity is often approximated by place-type similarity. In our case study, \textit{place} can be considered as a landmark (i.e., a geo-specific feature) within a larger geographic context, namely Vienna (which itself is also a \textit{place}). Consider the following famous landmarks in Vienna: St. Stephen's Cathedral is more similar to Karlskirche than to the Vienna State Opera house because the former two are churches, whereas the latter is not.

Place-type similarity can be computed using auxiliary knowledge bases, such as knowledge graphs~\citep{hogan2021knowledge}. Take Wikidata~\citep{vrandevcic2014wikidata}, one of the world's largest open knowledge graphs, as an example. Because Wikidata is a \textit{semantic network} of concepts, one can use the \textit{Rada distance}~\citep{rada1989development} to measure the shortest-path length between the nodes, i.e., concepts corresponding to Vienna's famous landmarks in Wikidata. The shortest-path length is equivalent to the minimum number of edges connecting them. Eq.\,(\ref{eq:rada}) presents the formula for the Rada distance, where $c_1$ and $c_2$ denote two concepts. 

\begin{equation}
\label{eq:rada}
d_{\text{Rada}}(c_1, c_2) = \textit{len}\big( \textit{shortest\_path}(c_1, c_2) \big)
\end{equation}

\begin{figure}[h!tbp]
    \centering
    \includegraphics[width=0.6\textwidth]{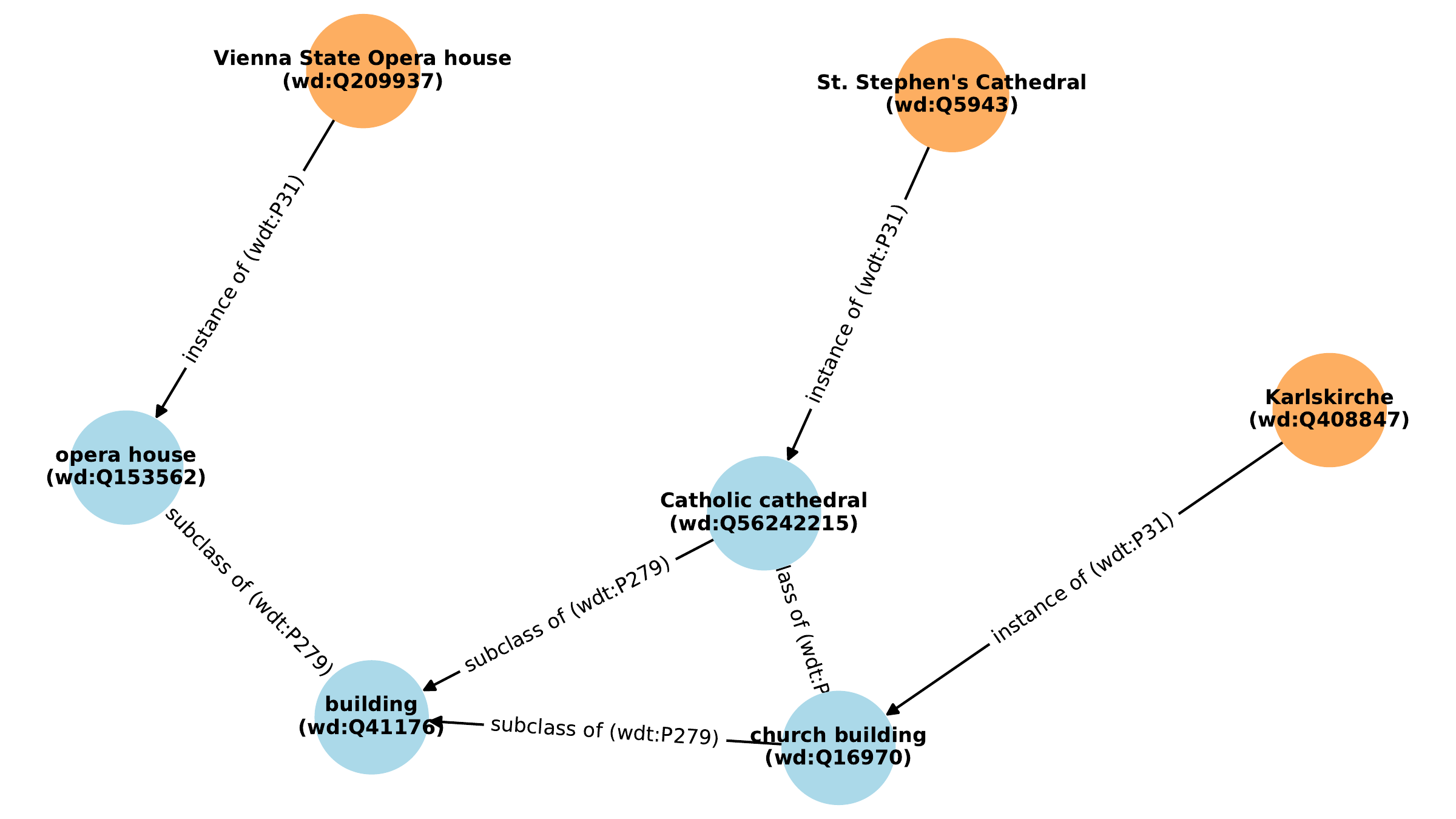}
    \caption{The shortest path between each pair of landmarks in the set \{St. Stephen's Cathedral, Karlskirche, Vienna State Opera house\} according to Wikidata. These landmarks are connected by the \textit{instance of} relation and its subsequent \textit{subclass of} relations. Blue nodes denote the landmarks. Orange nodes denote their Wikidata (super)classes, indicated when a blue node can reach an orange node through a directed edge. In Wikidata, each concept has a unique item identifier (e.g., wd:Q5943) and each relation has a unique property identifier (e.g., wdt:P31). These are illustrated here as well.}
    \label{fig:vienna_shortest_path_subgraph_full_iri_labeled}
\end{figure}

Fig.\,\ref{fig:vienna_shortest_path_subgraph_full_iri_labeled} shows a subgraph of Wikidata containing the three aforementioned landmarks. From this connected and directed graph, it can be derived that the Rada distance between St. Stephen's Cathedral and Karlskirche is 3, while it is 4 between the St. Stephen's Cathedral and the Vienna State Opera, as well as between Karlskirche and the Vienna State Opera.

The similarity between two nodes can then be measured as the reciprocal of the Rada distance, yielding a normalized measure where more closely connected nodes in a semantic network have higher similarity; see Eq.\,(\ref{eq:radasim}), where the addition of 1 in the denominator ensures that division by zero is avoided.

\begin{equation}
\label{eq:radasim}
\text{sim}_{\text{Rada}}(c_1, c_2) = \frac{1}{1 + d_{\text{Rada}}(c_1, c_2)}
\end{equation}

\subsubsection{Measurement Extension}
Eq.\,(\ref{eq:simdiv}) presents the similarity-sensitive diversity measure proposed by \citet{leinster2012measuring}, which we refer to as the \textit{Leinster–Cobbold number}. Here, $\mathbf{Z}=(z_{ij})$ is an $S \times S$ matrix, $S$ is the number of distinct places (or species), $p_i$ is the probability of a place randomly drawn from (Gen)AI outputs (or a species randomly drawn from an ecological community), and $z_{ij}$ denotes the similarity between the $i^{th}$ and $j^{th}$ places (or species). Specifically, $z_{ii}=1$. Note that as the similarity between places (or species) decreases, the Leinster-Cobbold number increases due to its monotonicity.

\begin{equation}
\label{eq:simdiv}
{}^q\!D^{\mathbf{Z}} =
\left(
\sum_{i=1}^{S} p_i 
\Bigg( \sum_{j=1}^{S} z_{ij} p_j \Bigg)^{q-1}
\right)^{\frac{1}{1-q}}
\end{equation}
When the similarity between different places (or species) is ignored, $\mathbf{Z}$ becomes an identity matrix. In this case, Eq.\,(\ref{eq:simdiv}) reduces to the classical Hill number given in Eq.\,(\ref{eq:div}). Note that the Hill number is always smaller than the Leinster-Cobbold number for the same value of $q$.

\begin{equation}
{}^q\!D = \left( \sum_{i=1}^{S} p_i^q \right)^{\frac{1}{1-q}}
\label{eq:div}
\end{equation}

\subsubsection{Measurement Details}
In our case study, we apply both the Hill number and the Leinster–Cobbold number to evaluate the geographic diversity of outputs from our selected models.

To support this, we manually identify the primary landmark in each generated image and in each revised prompt. For a generated image, we determine its primary landmark as the foreground landmark (e.g., St. Stephen's Cathedral in Fig.\,\ref{fig:multi_agent_example}). For a revised prompt, we determine its primary landmark as the first-mentioned landmark (e.g., St. Stephen's Cathedral, again, in Fig.\,\ref{fig:multi_agent_example}). Details of our landmark identification for the generated images are provided in Appendix~\ref{sec:landmark_identification_image}, while Appendix~\ref{sec:landmark_identification_prompt} describes the procedure for the revised prompts.

Next, we map the identified landmarks to their corresponding Wikidata entities. Details about the mapping results are provided in Appendix~\ref{sec:wikidata_rada}. Next, we extract their Wikidata property paths and compute the Rada distance to estimate place-type similarity. In our computation, we assume that all edges carry the same weight and that similarity is symmetrical.

\subsection{Data and Software Availability}

We store our data and codes in a GitHub repository\footnote{The repository can be accessed at \url{https://github.com/zilongliu-geo/image-gen-geodiversity}.}. The data includes the generated images and the revised prompts, along with the annotations. The code includes our data analysis and visualization.

\begin{figure*}[h]
    \centering
    \includegraphics[width=\textwidth]{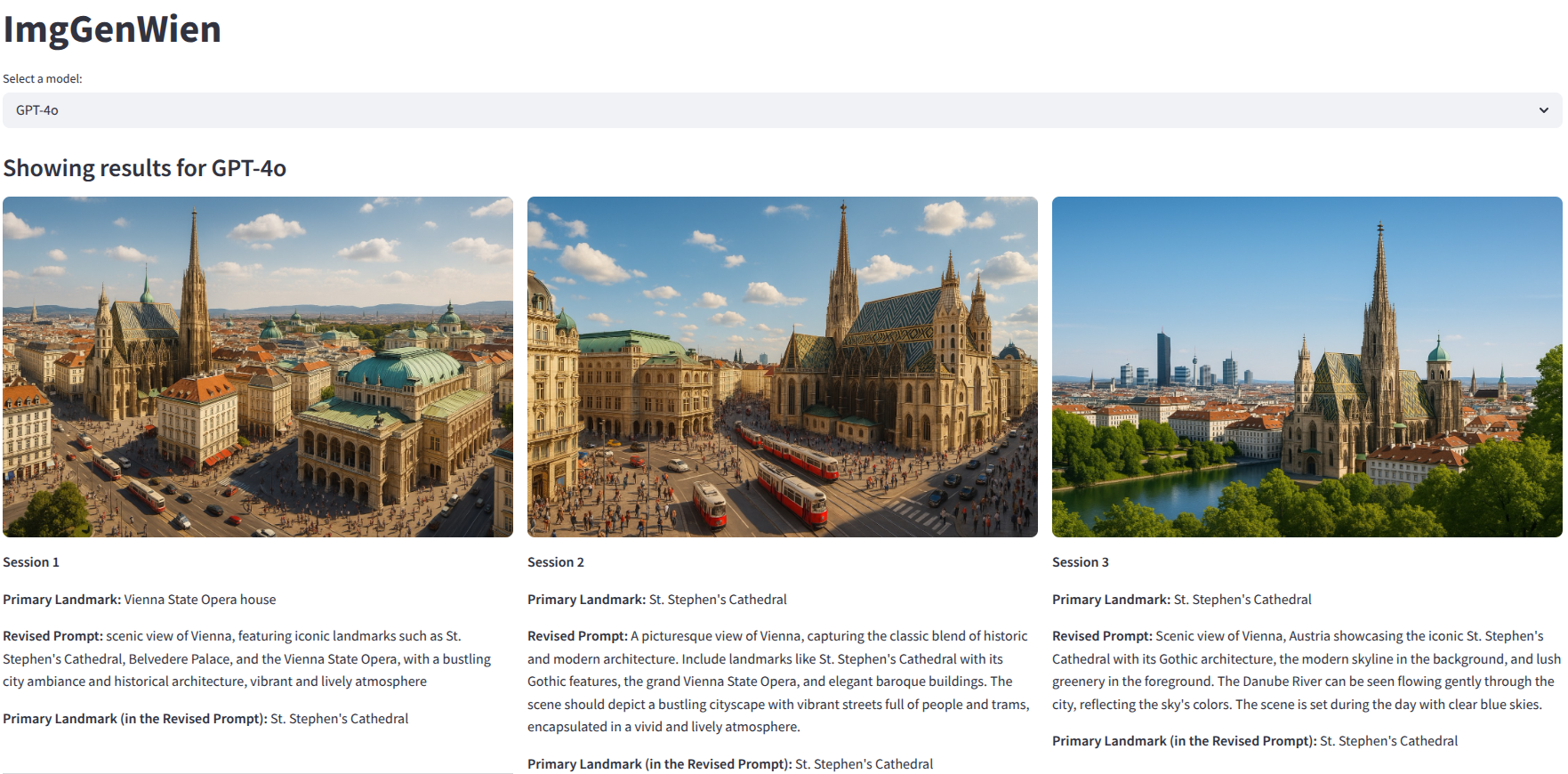}
    \caption{The observatory interface.  In this example, we select GPT-4o and are able to observe its generated images, automatically revised prompts, and annotations in the first three sessions in our experiment.}
    \label{fig:ImgGenWien}
\end{figure*}

To facilitate access to our annotations, we build a data application\footnote{The application can be accessed at \url{https://img-gen-wien.streamlit.app}.}. 
Fig.\,\ref{fig:ImgGenWien} illustrates its interface, showing how an OpenAI model depicts (or describes) Vienna.

\section{Results}
\label{sec:results}
In this section, we present our analysis results. These include \textbf{(1)} the proportion of \textit{valid} data among those generated by each model, \textbf{(2)} the degree to which an identified landmark is prototypical, \textbf{(3)} the similarity among our identified landmarks, and \textbf{(4)} the final diversity evaluation result.

First, the preliminary assessment on validity rates supports the diversity evaluation by distinguishing valid (which are those with identifiable landmarks) from invalid data, thereby determining their respective sampling probabilities during OpenAI's image generation. Second, the further assessment on landmark prototypicality informs the diversity evaluation by determining the cardinality $S$ of the outcome set $X$, as well as the probability distribution $p_i$ over the involved random variable $x$. Third, the landmark-similarity assessment provides the diversity evaluation with the similarity matrix $\mathbf{Z}$. Recall that these parameters are necessary for the computation of the Hill number and the Leinster-Cobbold number.

Notably, each assessment also yields surprising observations (as will be reported below).

\subsection{Validity Rates}
\begin{figure}[h!tbp]
    \centering
    \includegraphics[width=0.8\textwidth]{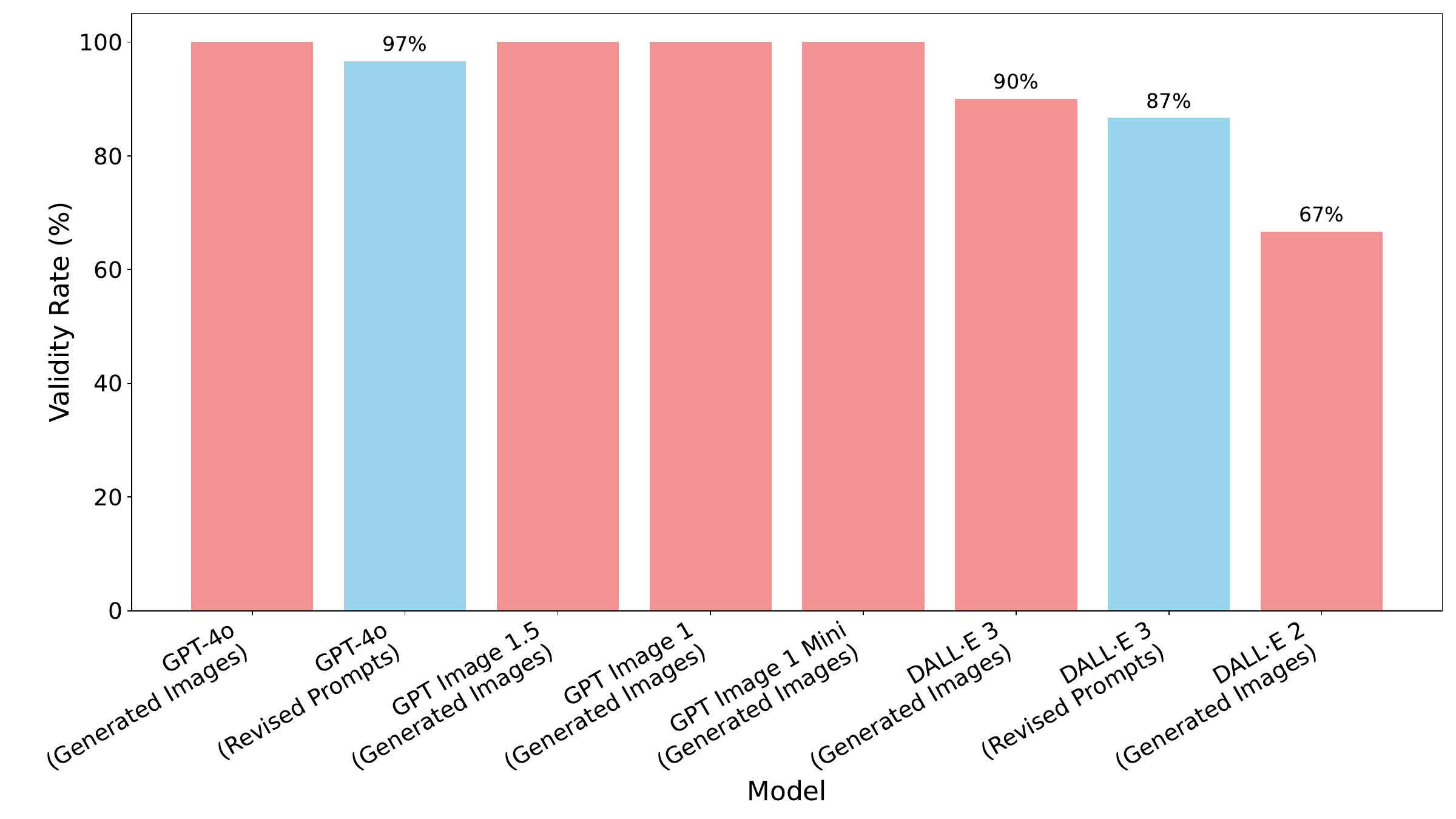}
    \caption{The validity rate of data generated by each selected model. Here, the rate is not shown if it reaches 100\%.}
    \label{fig:valid_images_per_model}
\end{figure}

Beyond evaluating diversity, it is also important to look at the proportion of valid images generated about Vienna, as a T2I model may not always be able to generate a human-interpretable image. Fig.\,\ref{fig:valid_images_per_model} illustrates the proportion of valid images generated by each model. DALL·E 2 has the lowest percentage (67\%), which is expected given that DALL·E 2 is the earliest OpenAI T2I model. DALL·E 3 performs substantially better, achieving 90\%. All other models, which are T2I models in the GPT family, reach 100\%. This indicates a significant improvement in the image-generation quality brought by OpenAI’s latest T2I series.

Additionally, Fig.\,\ref{fig:valid_images_per_model} displays the proportion of valid revised prompts (i.e., those containing identifiable landmarks) with respect to GPT-4o and DALL·E 3. Interestingly, compared with the generated images, the revised prompts have a lower validity rate for both models. This suggests that an OpenAI T2I model does not necessarily need linguistic clues to produce Vienna's visual descriptions containing its landmarks. We also observe that GPT-4o (97\%) produces more valid revised prompts than DALL·E 3 (87\%), although both rates are high. This indicates that when picturing Vienna, OpenAI's multi-agent system (based on GPT-4o) and DALL·E 3 both rely on linguistic clues, but GPT-4o does so to a greater extent.

\subsection{Landmark Prototypicality}
\begin{figure*}[h!tbp]
    \centering
    \includegraphics[width=\textwidth]{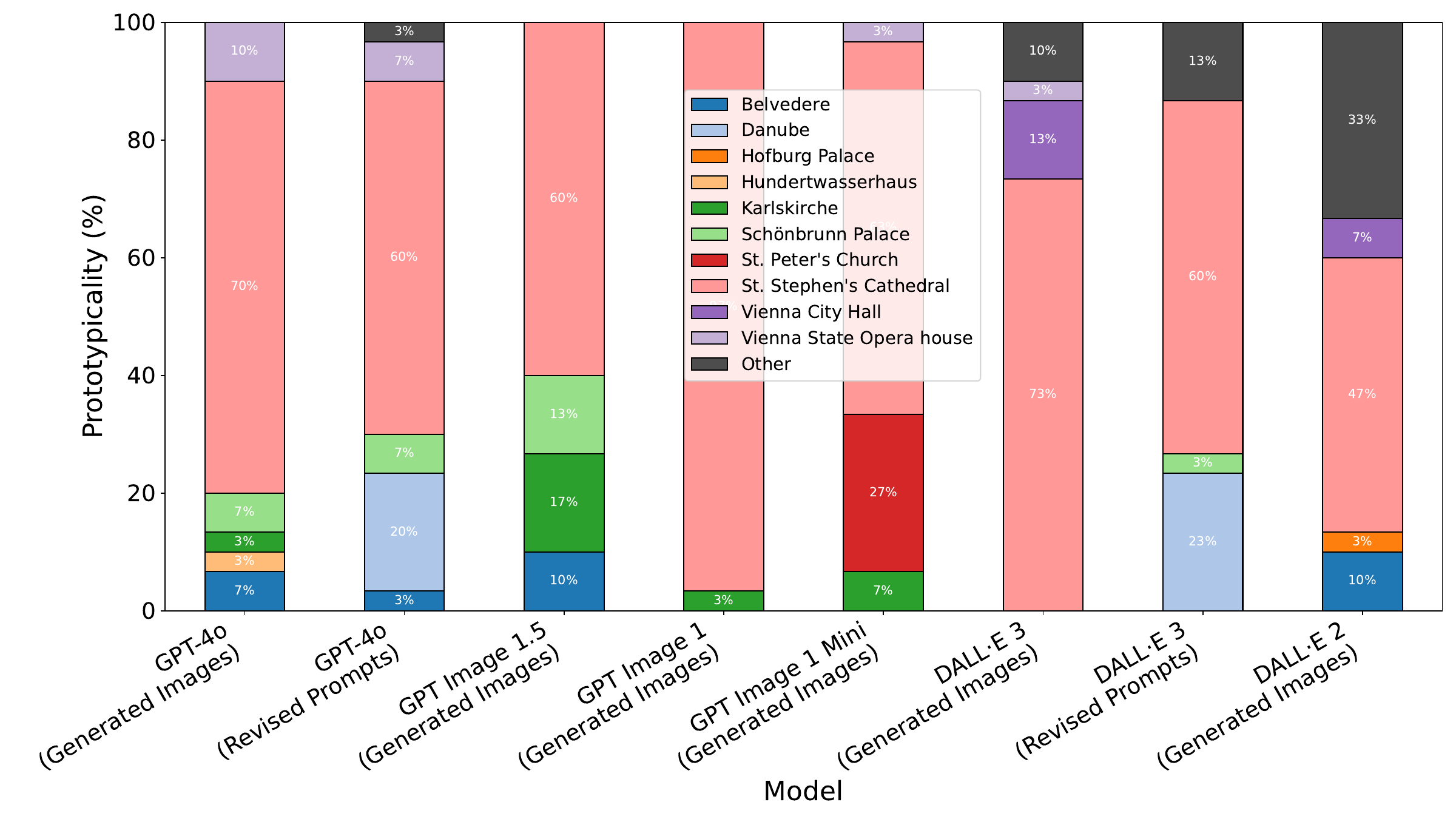}
    \caption{The prototypicality of primary landmarks identified in the data generated by each selected model.}
    \label{fig:landmark_distribution_per_model}
\end{figure*}

Initially, we hypothesize that the primary landmarks identifiable from the generated data will exhibit different proportions, thereby indicating varying degrees of prototypicality. Fig.\,\ref{fig:landmark_distribution_per_model} confirms this hypothesis by illustrating strong variations in landmark prototypicality, with St. Stephen's Cathedral appearing as the most prominent one. Here, the landmarks labeled as \textit{Other} correspond to invalid data.

When comparing our selected models, we observe that such prototypicality varies across them. For instance, GPT Image 1 exhibits the highest prototypicality (97\%) for St. Stephen's Cathedral, depicting only this landmark and Karlskirche. In contrast, Hofburg Palace is only present in the images generated by DALL·E 2 (3\%) and not in those produced by other models. Therefore, while the most prototypical landmark is shared by OpenAI’s image-generation models, the entire distribution of landmarks has notable variance, with those (e.g., much less prototypical landmarks) on the tail end of the distribution depicted (or mentioned) by only a few models.

We would like to approach this phenomenon from a \textit{Roschian} perspective, also known as the \textit{prototype theory}~\citep{rosch2024principles}. It states that a category can exhibit a graded structure, where certain members are more prototypical than others. Based on this cognitive-psychology theory, we consider that as evidenced by the varying prototypicality degrees, all our selected models (whether T2I models or LLMs) may possess a graded category structure about \textit{Vienna's landmarks}.

\subsection{Landmark Similarity}
\begin{figure*}[h!tbp]
    \centering
    \includegraphics[width=\textwidth]{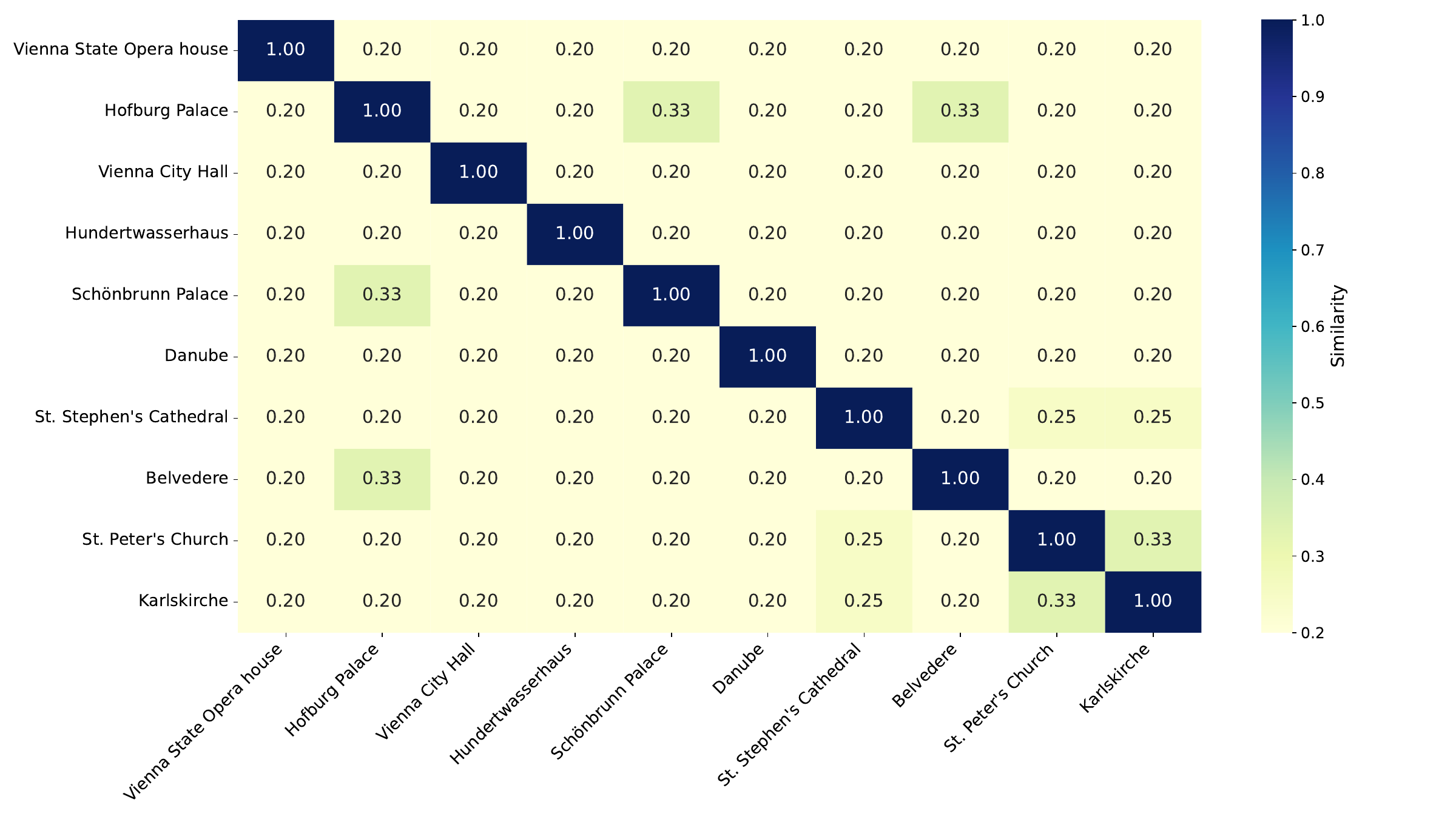}
    \caption{The similarity between each pair of identified landmarks.}
    \label{fig:wikidata_place_similarity_heatmap}
\end{figure*}

In our case study, landmark similarity is computed as a form of place-type similarity using Wikidata-based Rada distances. Fig.\,\ref{fig:wikidata_place_similarity_heatmap} illustrates the pairwise similarity between the identified landmarks. Aside from self-similarity (1.00), the highest landmark similarity is 0.33 for Hofburg Palace--Schönbrunn Palace, Hofburg Palace--Belvedere, and St. Peter's Church--Karlskirche. The second highest is 0.25 for St. Peter's Church--Karlskirche. All remaining similarity are 0.20, which is also the minimum similarity observed in the case study. 

\subsection{Diversity Profiles}
At this stage, all parameters required for our diversity measurement are prepared. The remaining question is how to visualize the results meaningfully. When \citet{leinster2012measuring} introduced species similarity into diversity measures, they also proposed a way for visualizing diversity, known as \textit{diversity profiles}. These are graphical representations that illustrate how diversity changes with respect to the order $q$. Here, we also adopt the use of diversity profiles to assess geographic diversity of the generated images and revised prompts.

\begin{figure*}[h!tbp]
    \centering
    \includegraphics[width=\textwidth]{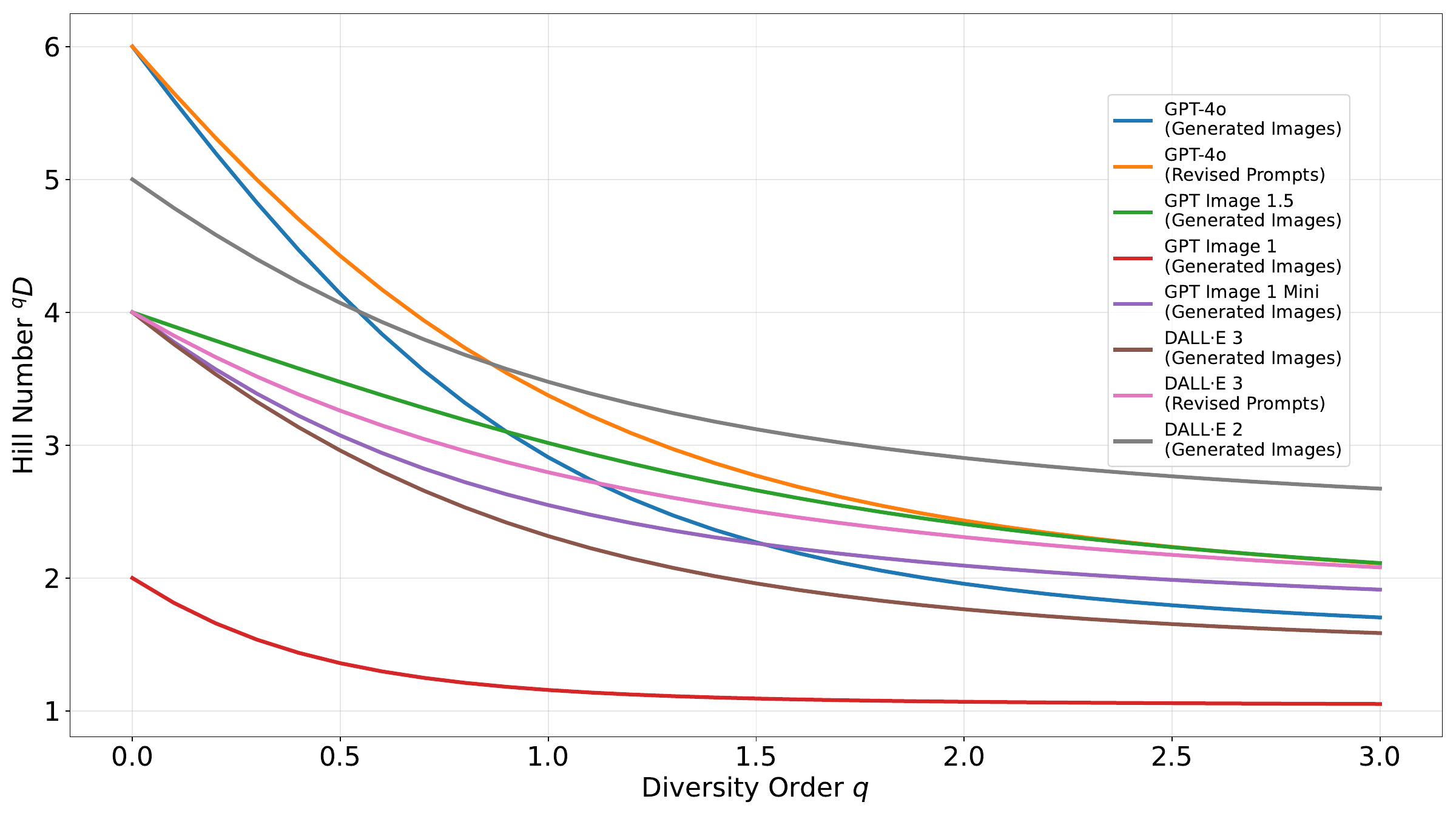}
    \caption{Diversity profiles of the generated data based on the Hill number.}
    \label{fig:diversity_profiles_hill_numbers}
\end{figure*}

\begin{figure*}[h!tbp]
    \centering
    \includegraphics[width=\textwidth]{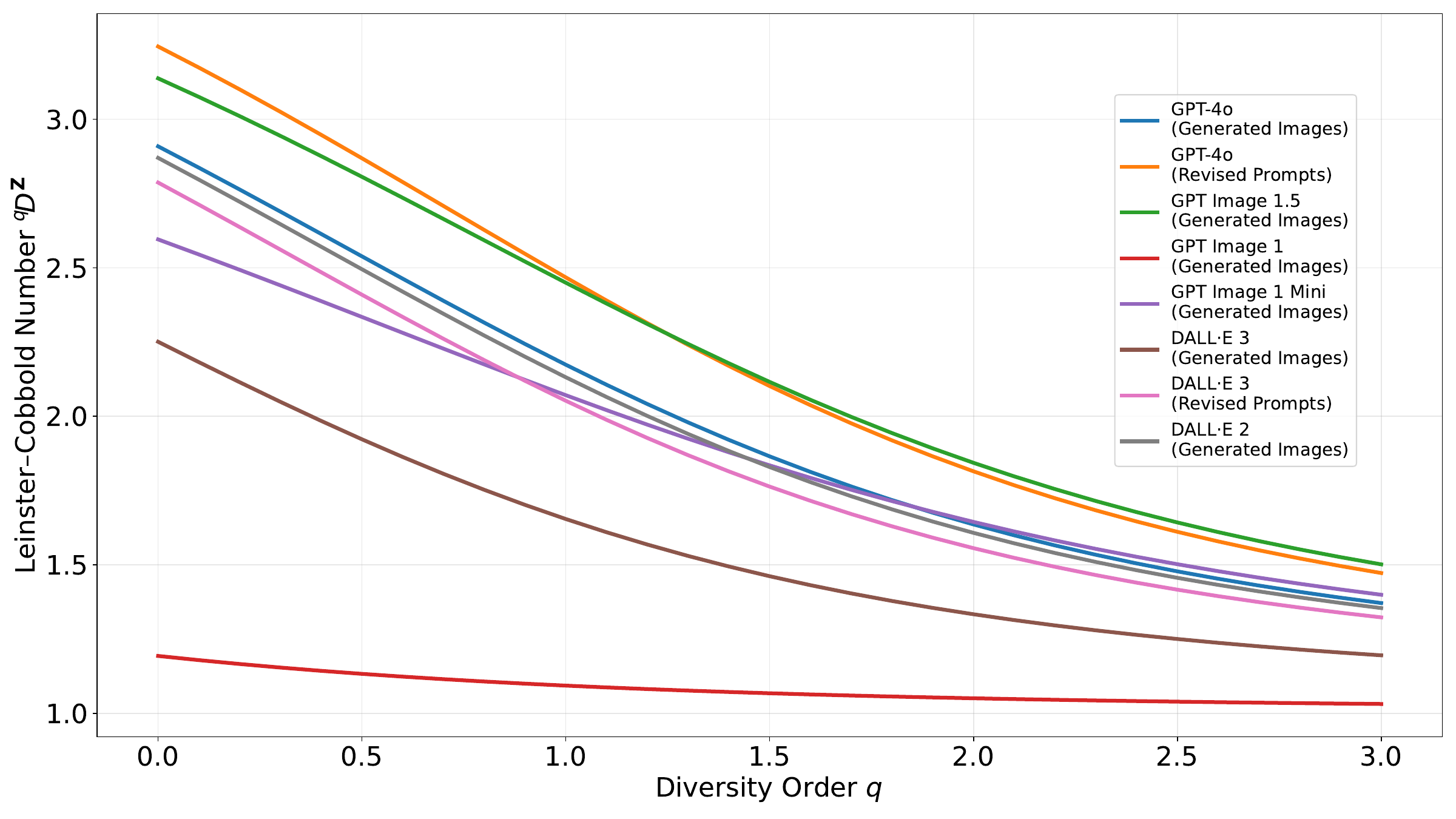}
    \caption{Diversity profiles of the generated data based on the Leinster-Cobbold number.}
    \label{fig:similarity_sensitive_diversity_all_models}
\end{figure*}

Fig.\,\ref{fig:diversity_profiles_hill_numbers} and Fig.\,\ref{fig:similarity_sensitive_diversity_all_models} illustrate the diversity profiles corresponding to the Hill number and the Leinster-Cobbold number, respectively. Focusing on where the two diversity profiles agree, we have two observations that are invariant regardless of $q$: \textbf{(1)} GPT Image 1 consistently exhibits the lowest diversity, and \textbf{(2)} DALL·E 3 consistently exhibits the second lowest diversity specifically in its image generation. Both observations suggest that a more recent T2I model does not necessarily exhibit higher diversity. This decline in geographic diversity, which occurs in spite of model development, is consistent with recent findings \citet{liu2025operationalizing} regarding autoregressive LLMs. 

As $q$ increases, common landmarks plays a bigger role in the resulting numbers. It can be, again, confirmed that none of our selected models depict a highly even distribution of landmarks, as their diversity profiles are all continuously decreasing curves. This is consistent with our earlier finding about the varying degrees of landmark prototypicality.

However, there are two other surprising observations. First, DALL·E 2 exhibits the third highest Hill number (at $q=0$) and later consistently the highest Hill number (roughly for $q>1$). This means that, when landmark similarity is not considered, DALL·E 2 (i.e., the oldest OpenAI T2I model) generates the most even distribution of landmarks. Second, GPT Image 1.5 exhibits the second highest Leinster-Cobbold number (roughly for $q<1$) and the highest Leinster-Cobbold number (roughly for $q>1.5$). This indicates that when landmark similarity is taken into account, GPT Image 1.5 (i.e., the newest OpenAI T2I model) depicts a relatively balanced distribution of landmarks. Both observations also highlight the necessity of accounting for similarity in diversity evaluation.

Moving towards cross-model comparision, we can observe that GPT Image 1 Mini consistently exhibits higher diversity than GPT Image 1, regardless of $q$ and whether similarity is considered. This suggests that GPT Image 1 Mini (i.e., a smaller version of GPT Image 1) outperforms its larger counterpart in terms of geographic diversity. Comparing prompt revision with image generation, we further observe that the revised prompts exhibit greater diversity than the generated images. This observation is supported by both DALL·E 3 and GPT-4o, again, regardless of $q$ and whether similarity is considered.

Together, these two figures suggest that for our case study, the Leinster-Cobbold number ($\text{maximum}<3.5$) is consistently smaller than the Hill number ($\text{maximum}=6$). We interpret this as a piece of geographical and empirical evidence that our selected models are considerably less diverse than what the Hill number naively reveals. In other words, our case study helps demonstrate that although the Hill number takes into account the prototypicality of each landmark, the Leinster-Cobbold number is a semantically more robust measure of diversity by additionally capturing the landmark similarity as a weighting factor.

\section{Discussion}
\label{sec:discussion}
At the outset of this paper, we have already argued that (Gen)AI's diversity issues have historic roots in the uncertainty and cognition of geographic information. Accordingly, we situate our discussion within these broader domains. We specifically link our methods and results to \textbf{(1)} uncertainty quantification and \textbf{(2)} environmental perception and cognition.

\subsection{Uncertainty Quantification}
In our work, diversity is considered an ethical concept. To encourage more research to engage in the observation of (Gen)AI, we propose looking into uncertainty as an information-theoretic form of diversity. As explained in Section~\ref{sec:background}, uncertainty can be quantified using (Shannon) entropy, and as such, it can be interpreted as \textit{the logarithm of diversity} from an information-theoretic perspective. This perspective also supports the study of (Gen)AI outputs through uncertainty quantification without invoking ethical considerations.

Several theoretical examples can complement this proposal. First, when \citet{leinster2012measuring} introduced their similarity-based extension to the Hill number, they also referenced a measure called \textit{Rao's quadratic entropy}~\citep{rao1982diversity}. Formally, it is a similarity-sensitive measure of uncertainty
. Second, similarity in a spatial sense is the foundation of \textit{Tobler's First Law of Geography}, which argues that ``near things are more related than distant things''~\citep{tobler1970computer}. Building on this principle, some GIScientists have conducted preliminary work on proposing entropy variants that account for spatial dependence~\citep{claramunt2005spatial}. From our perspective, Rao's quadratic entropy, Shannon entropy and its generalized versions (e.g., Rényi entropy~\citep{renyi1961measures}), spatially informed variants, and other entropy-based measures represent only a small subset of the extensive body of literature on uncertainty quantification.

\subsection{Environmental Perception and Cognition}
During our analysis about landmark prototypicality, we refer to prototype theory~\citep{rosch2024principles} as we interpret the uneven sampling probabilities as a manifestation of (cognitive) bias. While we do not equate (Gen)AI, which is a kind of non-biological entity, with humans, we also see our work relevant to the broader field of environmental perception and cognition, which examines how \textit{agents} psychologically respond to environments~\citep{garling1989environmental}.

It is worth noting that a growing body of literature is extending this field from humans to (Gen)AI, covering tasks such as predicting coordinates~\citep{bhandari2023large,roberts2023gpt4geo}, estimating distances~\citep{roberts2023gpt4geo}, and reasoning about qualitative spatial relations~\citep{bhandari2023large,fulman2024distortions,ji2025foundation}. However, most work has emphasized accuracy-oriented performance, focusing on how well a (Gen)AI makes predictions rather than on the variety and relative proportions of those predictions.

Hence, we propose looking into the diversity (or uncertainty) underlying the cognitive processing of (Gen)AI. This would require the integration of more sophisticated cognitive architectures. For instance, OpenAI models are now equipped with \textit{web-searching} capabilities~\citep{openai2025_web_search}, providing an interesting avenue for such investigations into the influence of \textit{retrieval-augmented generation}~\citep{lewis2020retrieval} on these processes.

\section{Conclusions}
\label{sec:conclusions}
In this work, we investigated the geographic diversity of (Gen)AI models used for image generation. We illustrated our methods and results using Vienna as a running example. By extending the Hill number, which is a similarity-insensitive diversity measure, we introduced the Leinster-Cobbold number. It helped incorporate categorical similarity among Vienna's landmarks. We then applied these measures to the diversity evaluation of text-to-image (T2I) generation models and relevant multimodal large language models (LLMs). Notably, considering that the outputs are multimodal, we examined not only the images generated by the DALL·E and GPT families of T2I models, but also the user prompts revised by DALL·E 3 and GPT-4o (i.e., an LLM).

We observed a strong prototypicality effect in the model outputs, which overwhelmingly centered on a small set of iconic landmarks. For instance, the selected models all tended to depict or describe St. Stephen’s Cathedral, i.e., the most central landmark in Vienna, while other landmarks appeared much less frequently (e.g., Vienna City Hall) or were entirely omitted (e.g., Musikverein, home of the world-renowned Vienna New Year's Concert). This reveals that these models represented Vienna as a narrowly defined place. However, this tendency is not surprising, as humans also commonly associate a city with its central landmarks and there is no lack of prominent landmarks in Vienna’s city center. A more important question is why these landmarks are selected over others, how the models justify their choices, and whether these justifications align with human reasoning.

In addition, more recent models, like GPT Image 1.5, did not necessarily exhibit higher geographic diversity. The same holds true for the newest multi-agent system based on GPT-4o and GPT Image 1 (Mini), even when landmark similarity was disregarded. Furthermore, the Leinster-Cobbold number consistently showed substantially lower geographic diversity than the Hill number, meaning that many so-called ``distinct” landmarks appearing in the generated data were, in fact, semantically close in terms of their landmark types. Therefore, the multimodal (Gen)AI ecosystem may be at risk of a (geographic) diversity collapse as it continues to develop.

Hence, we call for more systematic (Gen)AI diversity evaluations through place-centric case studies. Such case studies should span a range of geographic scales, look into geo-specific features, and incorporate experimental settings that vary in modalities, model-sampling parameters, and the inclusion of cognitive architectures. Future work could bring further insights from the extensive literature on uncertainty quantification as well as environmental perception and cognition, thereby reinforcing the uncertainty and cognitive pillars of GIScience.

\subsection*{Acknowledgements}
The authors would like to thank Songlin Wang and Annika Sü{\ss} for their helpful discussions.

\bibliographystyle{copernicus-agile}
\bibliography{example.bib}

\appendix
\section{Landmark Identification for the Generated Images}
\label{sec:landmark_identification_image}
During this object-detection process, we use Google Image Search\footnote{\url{https://images.google.com}} and Google Search\footnote{\url{https://www.google.com}} for reference. This is because the landmarks in certain images are difficult to recognize solely based on human vision.

We would like to list a few observations about the generated images along with how we address them. First, we observe that DALL·E 2 occasionally generates images that appear as mosaics of multiple sub-images. Because it is difficult to recognize landmarks in these sub-images, we do not assign any landmark in such cases. Second, DALL·E 3 generates hyper-realistic images. In certain cases, we resort to the \textit{AI mode} of Google Search to assist our landmark identification. Third, certain images generated by GPT Image 1 Mini include St. Peter's Church, a landmark geographically close to St. Stephen's Cathedral in reality. When St. Peter's Church is placed in the foreground ahead of St. Stephen's Cathedral, we record St. Peter's Church as the primary landmark, despite the greater global fame of St. Stephen's Cathedral.

Here, we attach Tables~\ref{tab:dalle2_landmarks},~\ref{tab:dalle3_landmarks},~\ref{tab:gpt_image_1_mini_landmarks},~\ref{tab:gpt_image_1_landmarks},~\ref{tab:gpt_image_1_5_landmarks}, and~\ref{tab:gpt_4o_landmarks}. These tables contain the landmarks identified in the generated images of our selected models. The \textit{Session} column indicates the session identifier associated with a generated image. The \textit{Note} column, when applicable, describes the content of an image when a primary landmark is not observed in its foreground.

At the time of writing, Google Search provides an \textit{AI Overview} for an uploaded image, which can sometimes return an automatically identified landmark based on its multimodal (Gen)AI system, i.e., Gemini\footnote{\url{https://gemini.google.com}}. This information is reported in the \textit{Note} column when available.
\begin{table}[t]
\centering
\caption{Identified landmarks in DALL·E 2's generated images.}
\label{tab:dalle2_landmarks}
\begin{tabular}{lll}
\toprule
 \textbf{Session} & \textbf{Landmark} & \textbf{Note} \\
\midrule
 1  & St. Stephen's Cathedral &  \\
 2  & / & Vienna's landmarks \\
 3  & / & Vienna's landmarks \\
 4  & Hofburg Palace &  \\
 5  & / & ``VIEEN'' and ``VUEN'' \\
 6  & / & Smolny Cathedral \\
 7  & / & ``VEERN'' \\
 8  & St. Stephen's Cathedral &  \\
 9  & / & Milan Cathedral \\
 10 & Belvedere &  \\
 11 & St. Stephen's Cathedral &  \\
 12 & / & Brussels Town Hall \\
 13 & Vienna City Hall &  \\
 14 & St. Stephen's Cathedral &  \\
 15 & St. Stephen's Cathedral &  \\
 16 & St. Stephen's Cathedral &  \\
 17 & Vienna City Hall &  \\
 18 & St. Stephen's Cathedral &  \\
 19 & St. Stephen's Cathedral &  \\
 20 & St. Stephen's Cathedral &  \\
 21 & / & Europe's landmarks\\
 22 & St. Stephen's Cathedral &  \\
 23 & St. Stephen's Cathedral &  \\
 24 & St. Stephen's Cathedral &  \\
 25 & / & Ulm Minster \\
 26 & Belvedere &  \\
 27 & / & Fisherman's Bastion \\
 28 & St. Stephen's Cathedral &  \\
 29 & Belvedere &  \\
 30 & St. Stephen's Cathedral &  \\
\bottomrule
\end{tabular}
\end{table}

\begin{table}[t]
\centering
\caption{Identified landmarks in DALL·E 3's generated images.}
\label{tab:dalle3_landmarks}
\begin{tabular}{lll}
\toprule
 \textbf{Session} & \textbf{Landmark} & \textbf{Note} \\
\midrule
 1& St. Stephen's Cathedral&  

\\
 2& St. Stephen's Cathedral& 

\\
 3& /& Vienna's landmarks
\\
 4& Vienna City Hall&  

\\
 5& St. Stephen's Cathedral& 

\\
 6& St. Stephen's Cathedral& 

\\
 7& St. Stephen's Cathedral& 

\\
 8& Vienna City Hall&  

\\
 9& St. Stephen's Cathedral& 

\\
 10& St. Stephen's Cathedral&  

\\
 11& St. Stephen's Cathedral&  

\\
 12& St. Stephen's Cathedral& 

\\
 13& St. Stephen's Cathedral&  

\\
 14& Vienna City Hall&  

\\
 15& Vienna State Opera house&  

\\
 16& Vienna City Hall&  

\\
 17& St. Stephen's Cathedral&  

\\
 18& St. Stephen's Cathedral&  

\\
 19& St. Stephen's Cathedral&  

\\
 20& St. Stephen's Cathedral&  

\\
 21& St. Stephen's Cathedral& 

\\
 22& St. Stephen's Cathedral&  

\\
 23& St. Stephen's Cathedral&  

\\
 24& St. Stephen's Cathedral&  

\\
 25& /& Austria's landmarks
\\
 26& St. Stephen's Cathedral&  

\\
 27& St. Stephen's Cathedral& 

\\
 28& St. Stephen's Cathedral&  

\\
 29& St. Stephen's Cathedral&  

\\
 30& /&  Dresden's landmarks
\\
\bottomrule
\end{tabular}
\end{table}

\begin{table}[t]
\centering
\caption{Identified landmarks in GPT Image 1 Mini's generated images.}
\label{tab:gpt_image_1_mini_landmarks}
\begin{tabular}{ll}
\toprule
 \textbf{Session} & \textbf{Landmark} \\
\midrule
 1& St. Stephen's Cathedral\\
 2& St. Stephen's Cathedral\\
 3& St. Stephen's Cathedral\\
 4& St. Stephen's Cathedral\\
 5& St. Stephen's Cathedral\\
 6& St. Stephen's Cathedral\\
 7& St. Stephen's Cathedral\\
 8& St. Stephen's Cathedral\\
 9& St. Stephen's Cathedral\\
 10& St. Stephen's Cathedral\\
 11& Karlskirche\\
 12& St. Peter's Church\\
 13& Vienna State Opera house\\
 14& Karlskirche\\
 15& St. Stephen's Cathedral\\
 16& St. Stephen's Cathedral\\
 17& St. Stephen's Cathedral\\
 18& St. Stephen's Cathedral\\
 19& St. Peter's Church\\
 20& St. Stephen's Cathedral\\
 21& St. Stephen's Cathedral\\
 22& St. Peter's Church\\
 23& St. Stephen's Cathedral\\
 24& St. Peter's Church\\
 25& St. Stephen's Cathedral\\
 26& St. Peter's Church\\
 27& St. Stephen's Cathedral\\
 28& St. Peter's Church\\
 29& St. Peter's Church\\
 30& St. Peter's Church\\
\bottomrule
\end{tabular}
\end{table}

\begin{table}[t]
\centering
\caption{Identified landmarks in GPT Image 1's generated images.}
\label{tab:gpt_image_1_landmarks}
\begin{tabular}{ll}
\toprule
 \textbf{Session} & \textbf{Landmark} \\
\midrule
 1& St. Stephen's Cathedral
\\
 2& St. Stephen's Cathedral
\\
 3& St. Stephen's Cathedral
\\
 4& St. Stephen's Cathedral
\\
 5& St. Stephen's Cathedral
\\
 6& St. Stephen's Cathedral
\\
 7& St. Stephen's Cathedral
\\
 8& St. Stephen's Cathedral
\\
 9& St. Stephen's Cathedral
\\
 10& St. Stephen's Cathedral
\\
 11& St. Stephen's Cathedral
\\
 12& St. Stephen's Cathedral
\\
 13& St. Stephen's Cathedral
\\
 14& Karlskirche
\\
 15& St. Stephen's Cathedral
\\
 16& St. Stephen's Cathedral
\\
 17& St. Stephen's Cathedral
\\
 18& St. Stephen's Cathedral
\\
 19& St. Stephen's Cathedral
\\
 20& St. Stephen's Cathedral
\\
 21& St. Stephen's Cathedral
\\
 22& St. Stephen's Cathedral
\\
 23& St. Stephen's Cathedral
\\
 24& St. Stephen's Cathedral
\\
 25& St. Stephen's Cathedral
\\
 26& St. Stephen's Cathedral
\\
 27& St. Stephen's Cathedral
\\
 28& St. Stephen's Cathedral
\\
 29& St. Stephen's Cathedral
\\
 30& St. Stephen's Cathedral
\\
\bottomrule
\end{tabular}
\end{table}
\begin{table}[t]
\centering
\caption{Identified landmarks in GPT Image 1.5's generated images.}
\label{tab:gpt_image_1_5_landmarks}
\begin{tabular}{ll}
\toprule
 \textbf{Session} & \textbf{Landmark} \\
\midrule
 1& St. Stephen's Cathedral
\\
 2& St. Stephen's Cathedral
\\
 3& St. Stephen's Cathedral
\\
 4& St. Stephen's Cathedral
\\
 5& St. Stephen's Cathedral
\\
 6& St. Stephen's Cathedral
\\
 7& Belvedere
\\
 8& St. Stephen's Cathedral
\\
 9& Karlskirche
\\
 10& Karlskirche
\\
 11& Karlskirche
\\
 12& St. Stephen's Cathedral
\\
 13& St. Stephen's Cathedral
\\
 14& Karlskirche
\\
 15& Schönbrunn Palace
\\
 16& St. Stephen's Cathedral
\\
 17& St. Stephen's Cathedral
\\
 18& Belvedere
\\
 19& St. Stephen's Cathedral
\\
 20& St. Stephen's Cathedral
\\
 21& Schönbrunn Palace
\\
 22& St. Stephen's Cathedral
\\
 23& St. Stephen's Cathedral
\\
 24& St. Stephen's Cathedral
\\
 25& Karlskirche
\\
 26& Belvedere
\\
 27& St. Stephen's Cathedral
\\
 28& St. Stephen's Cathedral
\\
 29& Schönbrunn Palace
\\
 30& Schönbrunn Palace
\\
\bottomrule
\end{tabular}
\end{table}
\begin{table}[t]
\centering
\caption{Identified landmarks in GPT-4o's generated images in collaboration with GPT Image 1 or GPT Image 1 Mini.}
\label{tab:gpt_4o_landmarks}
\begin{tabular}{ll}
\toprule
 \textbf{Session} & \textbf{Landmark} \\
\midrule
 1& Vienna State Opera house
\\
 2& St. Stephen's Cathedral
\\
 3& St. Stephen's Cathedral
\\
 4& St. Stephen's Cathedral
\\
 5& St. Stephen's Cathedral
\\
 6& St. Stephen's Cathedral
\\
 7& St. Stephen's Cathedral
\\
 8& St. Stephen's Cathedral
\\
 9& Karlskirche
\\
 10& Belvedere
\\
 11& St. Stephen's Cathedral
\\
 12& St. Stephen's Cathedral
\\
 13& St. Stephen's Cathedral
\\
 14& St. Stephen's Cathedral
\\
 15& St. Stephen's Cathedral
\\
 16& Schönbrunn Palace
\\
 17& Hundertwasserhaus
\\
 18& St. Stephen's Cathedral
\\
 19& St. Stephen's Cathedral
\\
 20& St. Stephen's Cathedral
\\
 21& St. Stephen's Cathedral
\\
 22& St. Stephen's Cathedral
\\
 23& Vienna State Opera house
\\
 24& Schönbrunn Palace
\\
 25& St. Stephen's Cathedral
\\
 26& St. Stephen's Cathedral
\\
 27& St. Stephen's Cathedral
\\
 28& Belvedere
\\
 29& St. Stephen's Cathedral
\\
 30& Vienna State Opera house
\\
\bottomrule
\end{tabular}
\end{table}

\section{Landmarks Identification for the Revised Prompts}
\label{sec:landmark_identification_prompt}

We identify the primary landmarks mentioned in the prompts revised by DALL·E 3 and GPT-4o. This toponym-recognition task, i.e., a sub-task of named entity recognition (NER), is performed by using spaCy\footnote{\url{https://spacy.io}} and its large pre-trained English pipeline, i.e., en\_core\_web\_lg.

Table~\ref{tab:spacy_place_categories} lists the spaCy's NER labels we use. We then manually check the recognition results to ensure accuracy.

\begin{table}[h]
\centering
\caption{The spaCy's built-in NER labels used in our toponym-recognition task.}
\label{tab:spacy_place_categories}
\begin{tabular}{l}
\hline
\textbf{SpaCy NER Label} \\
\hline
GPE (Geo-Political Entity)\\
FAC (Facility) \\
LOC (Location) \\
ORG (Organization) \\
\hline
\end{tabular}
\end{table}

In addition, we attach Table~\ref{tab:dalle3_prompts} and~\ref{tab:gpt_4o_prompts}. These tables contain the primary landmarks identified in the revised prompts of DALL·E 3 and GPT-4o. The \textit{Session} column indicates the session identifier associated with a revised prompt.

\begin{table}[t]
\centering
\caption{Identified landmarks in DALL·E 3's revised prompts.}
\label{tab:dalle3_prompts}
\begin{tabular}{ll}
\toprule
 \textbf{Session} & \textbf{Landmark} \\
\midrule
 1& St. Stephen's Cathedral
\\
 2& St. Stephen's Cathedral
\\
 3& St. Stephen's Cathedral
\\
 4& Danube
\\
 5& St. Stephen's Cathedral
\\
 6& /
\\
 7& St. Stephen's Cathedral
\\
 8& Schönbrunn Palace
\\
 9& Danube
\\
 10& St. Stephen's Cathedral
\\
 11& St. Stephen's Cathedral
\\
 12& /
\\
 13& St. Stephen's Cathedral
\\
 14& St. Stephen's Cathedral
\\
 15& Danube
\\
 16& Danube
\\
 17& St. Stephen's Cathedral
\\
 18& St. Stephen's Cathedral
\\
 19& St. Stephen's Cathedral
\\
 20& St. Stephen's Cathedral
\\
 21& Danube
\\
 22& St. Stephen's Cathedral
\\
 23& St. Stephen's Cathedral
\\
 24& /
\\
 25& Danube
\\
 26& St. Stephen's Cathedral
\\
 27& /
\\
 28& St. Stephen's Cathedral
\\
 29& St. Stephen's Cathedral
\\
 30& Danube
\\
\bottomrule
\end{tabular}
\end{table}
\begin{table}[t]
\centering
\caption{Identified landmarks in GPT-4o's revised prompts.}
\label{tab:gpt_4o_prompts}
\begin{tabular}{ll}
\toprule
 \textbf{Session} & \textbf{Landmark} \\
\midrule
 1& St. Stephen's Cathedral
\\
 2& St. Stephen's Cathedral
\\
 3& St. Stephen's Cathedral
\\
 4& St. Stephen's Cathedral
\\
 5& St. Stephen's Cathedral
\\
 6& Vienna State Opera house
\\
 7& St. Stephen's Cathedral
\\
 8& Vienna State Opera house
\\
 9& Danube
\\
 10& Belvedere
\\
 11& St. Stephen's Cathedral
\\
 12& /
\\
 13& St. Stephen's Cathedral
\\
 14& St. Stephen's Cathedral
\\
 15& St. Stephen's Cathedral
\\
 16& Schönbrunn Palace
\\
 17& St. Stephen's Cathedral
\\
 18& St. Stephen's Cathedral
\\
 19& St. Stephen's Cathedral
\\
 20& Danube
\\
 21& St. Stephen's Cathedral
\\
 22& Danube
\\
 23& Danube
\\
 24& St. Stephen's Cathedral
\\
 25& Danube
\\
 26& St. Stephen's Cathedral
\\
 27& St. Stephen's Cathedral
\\
 28& Schönbrunn Palace
\\
 29& Danube
\\
 30& St. Stephen's Cathedral
\\
\bottomrule
\end{tabular}
\end{table}

\section{Mapping Landmarks to Wikidata Items}
\label{sec:wikidata_rada}
We manually map each Vienna's primary landmark we identify to their corresponding Wikidata items. Table~\ref{tab:wikidata_landmarks} presents the mapping results.

\begin{table}[h]
\centering
\caption{Identified landmarks and their Wikidata item identifiers.}
\label{tab:wikidata_landmarks}
\begin{tabular}{ll}
\hline
\textbf{Landmark}&\textbf{Item Identifier}\\
\hline
Belvedere&Q211818\\
 Danube&Q1653\\
Hofburg Palace&Q46242\\
Hundertwasserhaus&Q493126\\
 Karlskirche&Q408847\\
 St. Peter's Church&Q693884\\
 Schönbrunn Palace&Q131330\\
 St. Stephen's Cathedral&Q5943\\
 Vienna State Opera house&Q209937\\
 Vienna City Hall&Q686468\\
 \hline
\end{tabular}
\end{table}

\end{document}